%% file: Fatayer_ExcitedStates_AFM2020.tex
\documentclass[
prl,reprint,
superscriptaddress,
showpacs,
amsmath,amssymb,
aps
]{revtex4-1}

\usepackage{graphicx}
\usepackage{multirow} 
\newcommand{\figref}[1]{Fig.~\ref{#1}}

\newcommand{\df}[0]{{\it $\Delta\text{f}$} }
\def\arrvline{\hfil\kern\arraycolsep\vline\kern-\arraycolsep\hfilneg}

\begin{document}

\title{Probing molecular excited states by atomic force microscopy}

\author{Shadi Fatayer}

\email[]{sfa@zurich.ibm.com}
\affiliation{IBM Research -- Zurich, S\"aumerstrasse 4, 8803 R\"uschlikon, Switzerland}

\author{Florian Albrecht}
\affiliation{IBM Research -- Zurich, S\"aumerstrasse 4, 8803 R\"uschlikon, Switzerland}
\author{Ivano Tavernelli}
\affiliation{IBM Research -- Zurich, S\"aumerstrasse 4, 8803 R\"uschlikon, Switzerland}
\author{Mats Persson}
\affiliation{Surface Science Research Centre, Department of Chemistry, University of Liverpool, Liverpool, L693BX, United Kingdom}

\author{Nikolaj Moll}
\author{Leo Gross}
\email[]{lgr@zurich.ibm.com}
\affiliation{IBM Research -- Zurich, S\"aumerstrasse 4, 8803 R\"uschlikon, Switzerland}

\date{\today}

\begin{abstract}
By employing single charge injections with an atomic force microscope, we investigated redox reactions of a molecule on a multilayer insulating film. First, we charged the molecule positively by attaching a single hole. Then we neutralized it by attaching an electron and observed three channels for the neutralization. We rationalize that the three channels correspond to transitions to the neutral ground state, to the lowest energy triplet excited states and to the lowest energy singlet excited states. By single-electron tunneling spectroscopy we measured the energy differences between the transitions obtaining triplet and singlet excited state energies. The experimental values are compared with density functional theory calculations of the excited state energies. Our results show that molecules in excited states can be prepared and that energies of optical gaps can be quantified by controlled single-charge injections. Our work demonstrates the access to, and provides insight into, ubiquitous electron-attachment processes related to excited-state transitions important in electron transfer and molecular optoelectronics phenomena on surfaces.

\end{abstract}


\maketitle

Scanning tunneling microscopy (STM) and spectroscopy (STS) are paramount to the electronic characterization of adsorbates on metallic and on ultrathin insulating films. Molecular resonances~\cite{Repp2005Pentacene}, molecular charging~\cite{WuPRL2004}, defect states~\cite{Repp2005Cl, NiliusPRL2003}, Kondo resonances~\cite{Vidya1998} and spin-flip excitations~\cite{Heinrich2004Science} are prominent examples of phenomena probed. STS experiments also complement other measurements, such as absorption spectroscopy~\cite{RauScience2014} and luminescence~\cite{Qiu2003Science, Imada2016, Doppagne2018, zhang2017sub, zhang2016,doppagne2020single}, providing further physicochemical insights.

Probing molecular charge transitions by STS in a double-barrier tunnel junction allows the mapping of densities of molecular frontier orbitals and measurement of the transport gap, also called fundamental gap, i.e. the difference between ionization potential and electron affinity~\cite{Repp2005Pentacene}. To this end, electrons/holes are attached with the STM tip. In a second step these excess charges tunnel between molecule and substrate reestablishing the charge ground state of the molecule. The second step is challenging to control in a conventional STM/STS experiment. By engineering the energy-level alignment of the surface work function with respect to the molecule's electron affinity and ionization potential, it is possible to obtain charge bistability~\cite{Leoni2011,Swart2011,Repp2006} or non-neutral charged ground states~\cite{kimura2019}. 

In STM, some limitations arise because typically only transitions whose initial state is the ground state can be probed. Multiple electron attachment to the ground state can be observed, as in multi-step excitations~\cite{Stipe1997,kim2002single} and up-conversion electroluminescence~\cite{chen2019spin}. In these cases, several charges are attached consecutively at one given bias voltage.

Some limitations can be overcome by the usage of multilayer insulating films (or bulk insulators~\cite{Rahe2016}) that suppress charge leakage to the metal substrate. On these films, out-of-equilibrium charge states can be prepared by charge injection from the tip~\cite{Steurer2015,berger2020quantum}. Moreover, the single-charge sensitivity of atomic force microscopy (AFM){\textbf~\cite{berger2020quantum,Steurer2015,Rahe2016,Stomp2005,Bussmann2006,Gross2009}} allows the measurement of charge transitions from the prepared out-of-equilibrium initial charge states. Several charges can be attached consecutively, each at a different bias voltage. This concept has recently been employed to measure the reorganization energy of a molecule~\cite{Fatayer2018}, to map densities of out-of-equilibrium charge transitions~\cite{Patera2018Nature} and to access charged intermediate states~\cite{patera2019prl}.

Here we show that after preparing a molecule in an out-of-equilibrium initial charge state by charge injection, we can probe the neutralization with a second charge injection as a function of voltage and create an excited neutral molecule. Specifically, we first attach a single hole to an individual naphthalocyanine (NPc) creating a cation (NPc$^+$) and then form a neutral molecule in the ground state or in excited states 
by electron attachment to NPc$^+$. Using single-electron AFM-based tunneling spectroscopy~\cite{Fatayer2018} we quantify the excitation energies. In addition, we calculated the excitation energies using time dependent density functional theory (TD-DFT) supporting our interpretation.  


The measurements were performed in a combined STM/AFM instrument equipped with a qPlus sensor \cite{Giessibl1998}. We used the frequency-modulation mode \cite{Albrecht1991} and an oscillation amplitude of 6\,\AA. The microscope was operated under ultrahigh vacuum ($p\,\approx\,10^{-11}\,\text{mbar}$) and low temperature ($T\,\approx\,5\,\text{K}$) conditions. Voltages were applied to the sample. As substrate we used a Cu(111) single crystal covered with NaCl layers of different thicknesses. One part of the crystal was pristine Cu(111) for tip preparation. The other part was covered with a multilayer NaCl film to prevent charge transfer to the metal substrate within the time-scales of the experiments~\cite{Steurer2014}. The experiments shown here were performed on a 14 monolayer thick film (see supplemental material for details). NPc molecules were deposited onto the cold (10 K) sample. Cu terminated AFM tips were prepared by controlled indentation into the pristine copper surface.

\begin{figure}
\includegraphics[width=0.6\columnwidth]{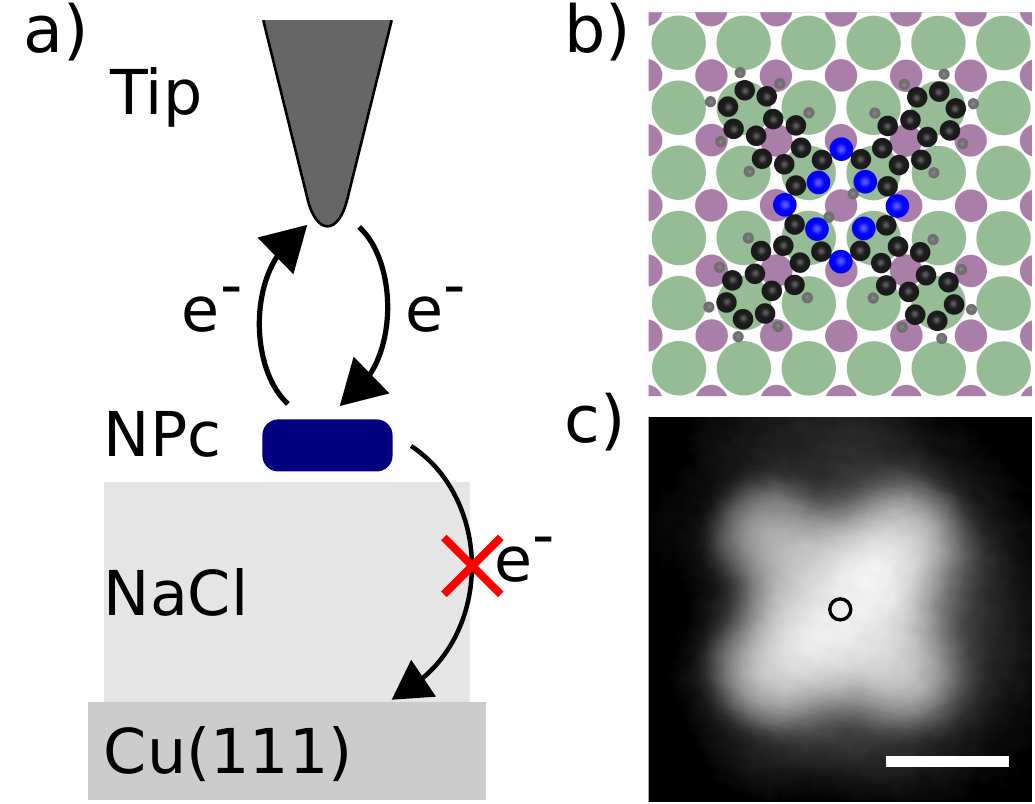}
\caption{\label{fig:Intro}{\bf(a)} Sketch of the experimental arrangement. The molecule (NPc) is adsorbed on top of a multilayer insulating film. Electron transfer is only possible between the tip and NPc, not between NPc and the metal substrate, Cu(111). {\bf(b)} Top view of the atomic model of NPc on top of NaCl. Blue, black and grey solid circles represent N, C and H elements, respectively. Purple and green solid circles represent the ions in the film, Na and Cl, respectively. {\bf(c)} Constant-\df AFM image (\df = -1.7\,Hz and {\it V}~=~1\,V) of NPc on multilayer NaCl. The scale bar is 10 \AA. Black to white color-scale corresponds to 3\AA~tip-height range. The black circle represents the position where the tunneling spectroscopy experiments were performed.}
\end{figure}

By employing a 14 monolayer insulating NaCl film, the only possible electron-transfer pathway is between the tip and NPc, as sketched in \figref{fig:Intro}(a). NPc, whose atomic model is displayed in \figref{fig:Intro}(b), is imaged on the multilayer NaCl film by AFM with a metal tip, in constant-\df mode. The AFM image reveals the expected cross shape of NPc, see \figref{fig:Intro}(c). 

\begin{figure*}
	\includegraphics[width=1.2\columnwidth]{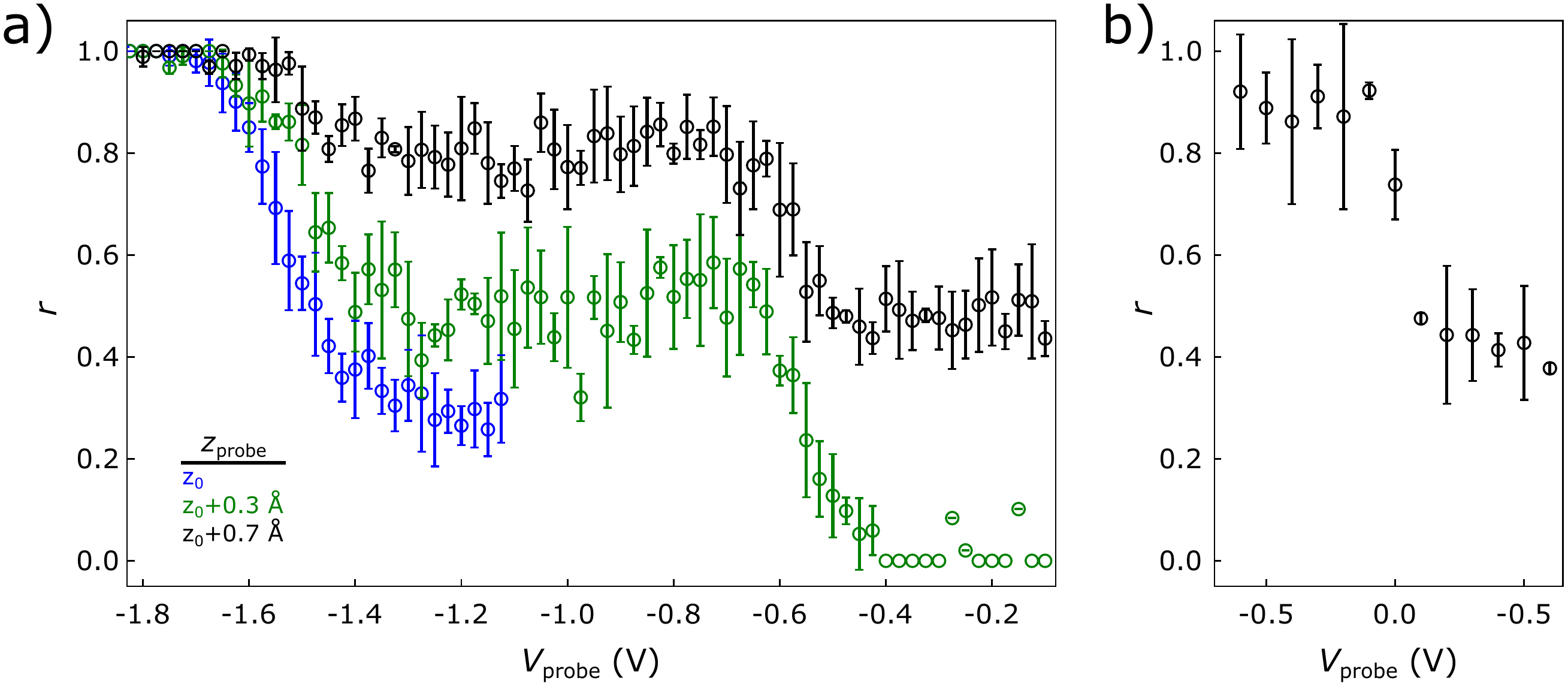}
	\caption{{\bf(a)} Fraction $r$ (time in the initial positive charge state divided by the probe time) as a function of {\it V}$_{\text{probe}}$ for different {\it z}$_{\text{probe}}$ for  NPc on a 14 ML NaCl. Each data point is extracted from 80 $\Delta f(t)$ probe traces. The error bars correspond to the standard error of the mean $r$ for each individual set of measurements of $r$. The data for {\it z}$_{\text{probe}}$ = z$_0$ is reproduced from~\cite{Fatayer2018}. {\bf(b)} Same as {\bf(a)}, with a different tip at larger {\it V}$_{\text{probe}}$. Each data point is extracted from 40 $\Delta f(t)$ probe traces.  
	}
	\label{fig:r_evolution}
\end{figure*} 

\begin{figure*}
	\includegraphics[width=1.2\columnwidth]{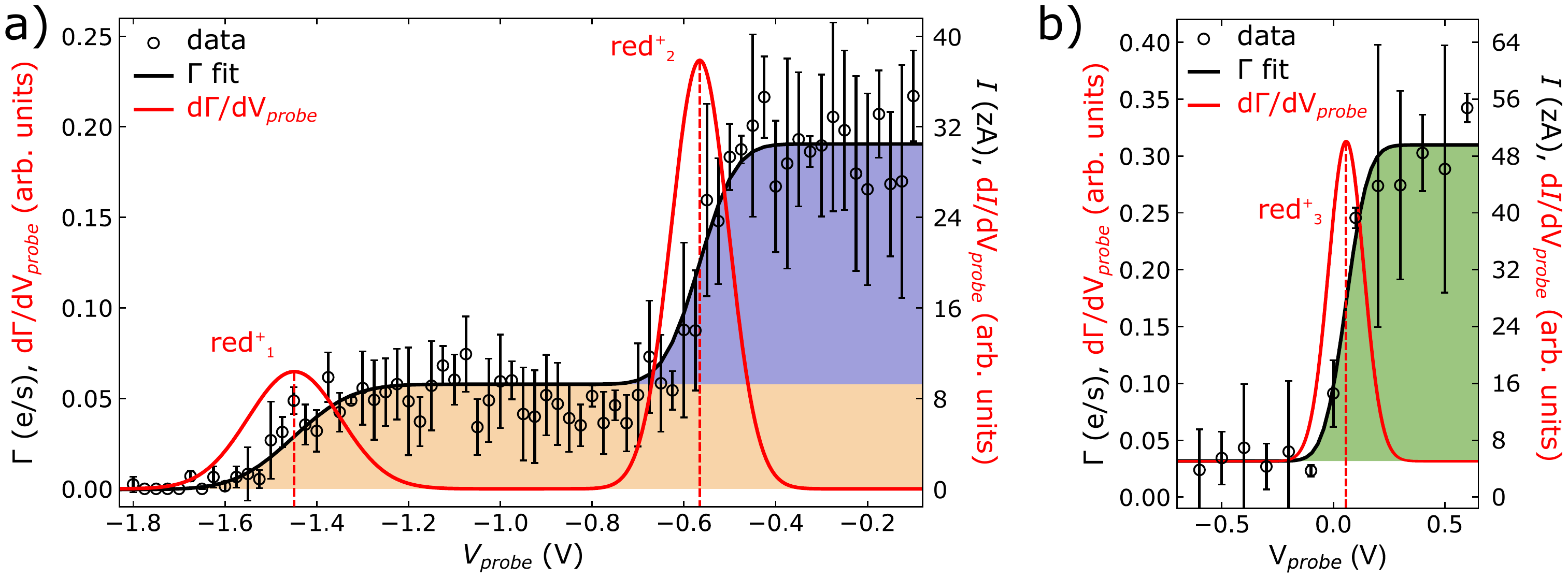}
	\caption{\label{fig:tunnel_spectroscopy}(Color online) {\bf(a)} Analysis of the $\text{NPc}^{+}\rightarrow\text{NPc}^{0}$ transition as a function of {\it V}$_{\textit{probe}}$ at {\it z}$_{\textit{probe}}$ = z$_0$+0.7\,\AA. Black circles represent the extracted tunneling rate $\Gamma$ (the corresponding value of tunneling current $I$ is displayed on the right axis) along with  the fitted sum of two error functions (black line) and its derivative (red line).
	The peak positions of red$^{+}$$_{1}$ and red$^{+}$$_{2}$ (dashed vertical lines) are indicated in the graph. The areas corresponding to each fitted error function are colored orange for the red$^{+}$$_{1}$ and purple for the red$^{+}$$_{2}$ transition, respectively. {\bf(b)} Same analysis as in {\bf(a)} but for larger {\it V}$_{\textit{probe}}$ and a different tip. The peak position of red$^{+}$$_{3}$ (dashed vertical line) is indicated in the graph. The area corresponding to the fitted error function is colored green.}
\end{figure*}
The spectroscopic investigation of the $\text{NPc}^{+}\rightarrow\text{NPc}^{0}$ transition  requires a recently introduced method that uses AFM to detect single-electron tunneling events between the tip and NPc~\cite{Fatayer2018}. Succinctly, we first set the molecule in the desired initial charge state, in this case NPc$^{+}$, by applying a voltage significantly below the oxidation voltage of NPc ({\it V} $<$ -3.2\,V). Subsequently, the electron-transfer is probed at appropriate voltage values {\it V}$_{\textit{probe}}$ and at fixed tip-molecule distance {\it z}$_{\text{probe}}$. Changes in the charge state of the molecule are identified as steps in the frequency-shift signal as a function of time $\Delta f(t)$. The statistical variation in the single-electron tunneling process is accounted for by recording 80 $\Delta f(t)$ probe traces in total (unless specified otherwise) at a fixed {\it V}$_{\textit{probe}}$ . 

\figref{fig:r_evolution} shows the fraction $r$, which is the averaged time of the molecule in the initial (positive) charge state divided by the probe time ({\it T}$_{\textit{probe}}$ $\approx$ 9\,s), when probing the transition from NPc$^+$ to NPc$^0$ at a fixed {\it V}$_{\textit{probe}}$ and {\it z}$_{\textit{probe}}$. The lifetime $T$ and its inverse, the tunneling rate $\Gamma$, can be extracted from $r$ (see supplemental material). The necessary condition for resolving the lifetime $T$ is that $T$ should be within the dynamic range of our experiment: if $T << 0.1$\,s (AFM time resolution), $r$ = 0. If $T >> 9$\,s ({\it T}$_{\textit{probe}}$), $r$ = 1. The tip-molecule distance {\it z}$_{\textit{probe}}$ is the parameter that allows adjusting $T$ to fit within the dynamic range of our experiment.

Different transitions can be accessed with different {\it z}$_{\textit{probe}}$. At {\it z}$_{\textit{probe}}$ = z$_0$, corresponding to a separation of (21 $\pm $ 5)\,\AA~between tip apex and molecule, we previously measured the transition red$^{+}$$_{1}$ at $\approx$~-1.5\,V~\cite{Fatayer2018}, assigned as a reduction of the cation, see the blue data in \figref{fig:r_evolution}(a). Increasing {\it z}$_{\textit{probe}}$ results in a decrease in $T$ and an increase in $r$. The {\it z}$_{\textit{probe}}$ increase causes an upward shift of the voltage window at which we are sensitive, given our dynamic range. At {\it z}$_{\textit{probe}}$ = z$_0$+0.3\,\AA, and with an increased {\it V}$_{\textit{probe}}$ range, we observe a second transition, red$^{+}$$_{2}$, at $\approx$ -0.6\,V, see green data in \figref{fig:r_evolution}(a). To quantify this transition, we further increased {\it z}$_{\textit{probe}}$ to z$_0$+0.7\,\AA, see black data in \figref{fig:r_evolution}(a).

With a different tip and adjusting {\it z}$_{\textit{probe}}$ to be sensitive at even larger {\it V}$_{\textit{probe}}$, we observed a third transition, red$^{+}$$_{3}$, see \figref{fig:r_evolution}(b). In addition to our current-meter dynamic range, there is a second condition that needs to be fulfilled to resolve a transition: The Kelvin probe force spectroscopy parabolas of NPc$^0$ and NPc$^+$ must not cross in the voltage range of the transition, as otherwise the initial and final state cannot be discerned in $\Delta f(t)$ probe traces. For that reason, the red$^{+}$$_{3}$ transition was measured with a different tip (supplemental material).

\figref{fig:tunnel_spectroscopy}(a) and (b) show the extracted tunneling rates $\Gamma$ corresponding to the black data points in \figref{fig:r_evolution}(a) and the data of \figref{fig:r_evolution}(b), respectively. The differential conductance related to charge-state transitions by resonant tunneling on NaCl is well described by a Gaussian~\cite{Repp2005Cl,Repp2005Pentacene}. The broadening of the peaks originates mainly from electron-phonon and electron-vibron coupling~\cite{Repp2005Cl,pavlivcek2013symmetry}. 
The evolution of $\Gamma$({\it V}$_{\textit{probe}}$) is well described by two error functions (black line indicates the sum of both error function fits) in \figref{fig:tunnel_spectroscopy}(a), and by one error function in \figref{fig:tunnel_spectroscopy}(b). The colored areas represent the contribution of each transition (via an error function) to the overall fit. 
The determined voltages of the resonances (dashed red lines in \figref{fig:tunnel_spectroscopy}) are {\it V}(red$^{+}$$_{1}$) = (-1.45 $\pm $ 0.05)\,V, {\it V}(red$^{+}$$_{2}$) = (-0.57 $\pm $ 0.04)\,V and {\it V}(red$^{+}$$_{3}$) = (+0.06 $\pm $ 0.07)\,V. The errors stem mainly from the uncertainty in determining the plateaus of $\Gamma$({\it V}$_{\textit{probe}}$).  
The partial voltage drop across the 14 ML NaCl dielectric film was estimated as $\approx$ 17~$\%$~\cite{Fatayer2018}. The energies corrected for the voltage drop across the NaCl are $E($red$^{+}$$_{1})=$ (-1.20 $\pm $ 0.15)\,eV, $E($red$^{+}$$_{2})=$(-0.47 $\pm $ 0.10)\,eV and $E($red$^{+}$$_{3})=$ ($+$0.05 $\pm $ 0.18)\,eV, respectively. The full-width at half-maximum (FWHM) of the red$^{+}$$_{1}$, red$^{+}$$_{2}$ and red$^{+}$$_{3}$ resonances are (0.23 $\pm $ 0.10)\,V, (0.15 $\pm $ 0.10)\,V and (0.18 $\pm $ 0.12)\,V, respectively. 

The peak position of the red$^{+}$$_{1}$ transition is the same as previously reported~\cite{Fatayer2018}. 
The Kelvin parabolas extracted from the $\Delta f(t)$ spectra (see supplemental material) evidence that red$^{+}$$_{1}$, red$^{+}$$_{2}$ and red$^{+}$$_{3}$ are charge transitions between NPc$^{+}$ and NPc$^{0}$. The $\text{NPc}^{0}\rightarrow\text{NPc}^{-}$ resonance is at {\it V}$_{\text{probe}}$ = +1.3\,V~(see supplemental material). 

\begin{figure*}
	\includegraphics[width=\linewidth]{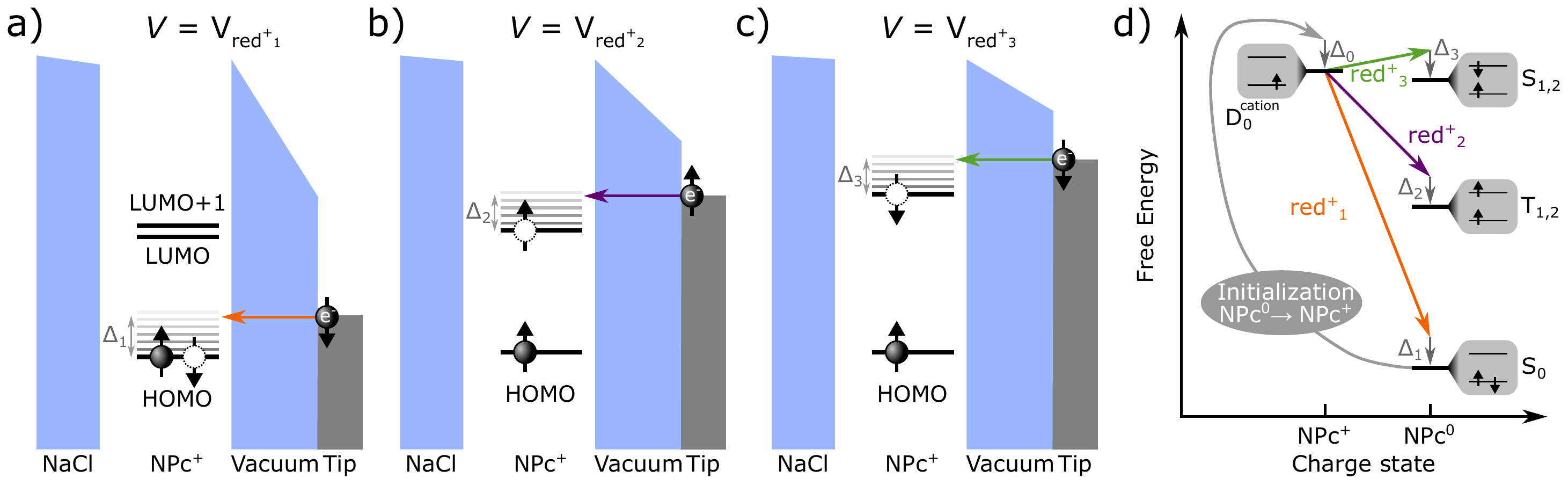}\caption{\label{fig:mechanism}(Color online) Diagrams of the single-electron transfer from the tip to the cation, resulting in NPc$^0$ in the {\bf(a)} ground state S$_0$, {\bf(b)} lowest energy triplet excited states T$_{1,2}$ and {\bf(c)} lowest energy singlet excited states S$_{1,2}$, respectively. Labels LUMO+1, LUMO and HOMO correspond to the second lowest unoccupied, lowest unoccupied and highest occupied molecular orbital, respectively, of the neutral molecule. Because of their small energy separation, LUMO and LUMO+1 are represented as one energy level in panels (b-d). The relaxation energies related to the respective charge transitions are labeled $\Delta_1$, $\Delta_2$ and $\Delta_3$. {\bf(d)} Many-body electron picture associated with the measured charge-state transitions (orange, purple and green). The controlled, initial cation D$_0^{\text{cation}}$ formation is highlighted with a grey arrow. The electronic structure of each state is shown in the grey insets. The relaxation energy related to the electron detachment from NPc is labeled $\Delta_0$.}
\end{figure*}

NPc$^{0}$ has a well-defined highest-occupied molecular orbital (HOMO) and lowest unoccupied molecular orbitals LUMO and LUMO+1 that are very close in energy~\cite{liljeroth2007}. For the rest of the text, HOMO, LUMO and LUMO+1 will refer to the frontier orbitals of NPc$^{0}$. Experimentally, we initially removed an electron from the HOMO, so that the molecule became positively charged, and its neutralization could then be probed via AFM-based tunneling spectroscopy. The NPc$^{+}$ highest-occupied state has a single-electron occupation. The transition red$^{+}$$_{1}$ can then be understood as the attachment of an electron, from the tip, directly into this singly-occupied molecular orbital~\cite{liljeroth2007} and reestablishing NPc$^{0}$ in the S$_0$ ground state. This process is depicted in \figref{fig:mechanism}(a). Since red$^{+}$$_{2}$ and red$^{+}$$_{3}$ occur at greater sample voltages than red$^{+}$$_{1}$, we conjecture that the electron is attached to the nearest-energy unoccupied states of the molecule resulting in the creation of excited states. The excited states that correspond to electron attachment to the LUMO and LUMO+1, forming a triplet excited state are T$_1$ and T$_2$. Another possibility is the formation of singlet excited states S$_1$ and S$_2$. 
The similar energies of LUMO and LUMO+1 in NPc and in phthalocyanines result in similar energies of T$_1$ and T$_2$, and of S$_1$ and S$_2$~\cite{doppagne2020single,Imada2016}. We assume that with our energy resolution we cannot distinguish between T$_1$ and T$_2$ (and neither between S$_1$ and S$_2$). Typically the lowest triplet excitation is at significantly smaller energies than the lowest singlet excitation.
Therefore, we assume that red$^{+}$$_{2}$ and red$^{+}$$_{3}$ correspond to transitions to the lowest energy triplet excited states T$_{1,2}$ and the lowest energy singlet excited states S$_{1,2}$, respectively, see \figref{fig:mechanism}(b,c). The transitions are sketched in a many-electron energy level diagram in \figref{fig:mechanism}(d). Calculating the energy difference between the measured transitions we obtain 
$E($red$^{+}$$_{2})$~-~$E($red$^{+}$$_{1})$~=~(0.73 $\pm $ 0.18)\,eV, which we assign to the triplet excited state energy (T$_{1,2}$~-~S$_0$)$^{\text{exp}}$, and 
$E($red$^{+}$$_{3})$~-~$E($red$^{+}$$_{1})$=~(1.25 $\pm $ 0.23)\,eV, which we assign to the singlet excited state energy (S$_{1,2}$~-~S$_0$)$^{\text{exp}}$.  

TD-DFT was employed to help clarify our experimental observations. The calculations were performed using the CPMD~\cite{car1985unified} code (in the supplemental material we also compared the results with the FHI-AIMS~\cite{blum_ab_2009} code). The geometries of the neutral and positively charged molecule were optimized with the {\it tight} basis defaults in the gas-phase and on top of a 2 layer NaCl slab with 100 atoms per layer. For structural relaxation, the Perdew-Burke-Ernzerhof (PBE) exchange-correlation functional was applied~\cite{perdew_generalized_1996} with vdW correction~\cite{tkatchenko_accurate_2009}. A planewave cutoff of 70 Ry at neutral global charge was used. 
More details are provided in the supplemental material. For NPc in vacuum, and in the geometry of the neutral molecule, the calculated lowest triplet excited state energy  is (T$_1$-S$_0$)$_{\text{vac}}^{\text{th}}$ = 0.95\,eV and its lowest singlet excited state energy is (S$_1$-S$_0$)$_{\text{vac}}^{\text{th}}$ = 2.10\,eV. For NPc adsorbed on NaCl, we obtained (T$_1$-S$_0$)$_{\text{NaCl}}^{\text{th}}$ = 0.90\,eV and (S$_1$-S$_0$)$_{\text{NaCl}}^{\text{th}}$ = 1.67\,eV. The values on NaCl are in good (T$_1$-S$_0$) and reasonable (S$_1$-S$_0$) agreement to the experiment. They support our assignment of the transitions confirming that the lowest energy triplet excited states are at smaller energies than the lowest energy singlet excited states. 
In comparison to the calculations, the systematic underestimation of the experimental values can be attributed in part to the negative differential resistance (NDR) observed at voltages larger than electron attachment resonances~\cite{Repp2005Pentacene}. Moreover, because of the NDR effect we are presumably more sensitive to the T$_1$ and S$_1$ transitions than to the T$_2$ and S$_2$ transitions, as the latter will be at voltages of NDR resulting from the former. Performing a more elaborate theoretical benchmarking with several functionals~\cite{maurer2019advances} is beyond the scope of our work. 

Via optical spectroscopy in solution the fluorescence (S$_{1,2}$~$\rightarrow$~S$_0$ transition) of NPc was measured at 801 nm (1.55 eV)~\cite{kumar2003spectral}. Its phosphorescence (T$_{1,2}$~$\rightarrow$~S$_0$ transition) has not been determined before.  Yet, for Silicon naphthalocyanine, the phosphorescence  was measured at 1330 nm (0.93 eV) and the fluorescence at 774 nm (1.60 eV)~\cite{firey1988silicon}. These literature values further support our assignment of the transitions.  


Note that in our experiment the redox transitions are measured in the NPc$^+$ geometry. With DFT we find that the excitation energies calculated in the NPc$^+$ geometry are about~0.05\,eV larger than the excitation energies calculated in the NPc$^0$ geometry (see supplemental material).


By analyzing the plateaus in the extracted tunneling current $I$ in \figref{fig:tunnel_spectroscopy}(a), $I$ increases by a factor of $\approx$~3 at the transition red$^{+}$$_{2}$, and $I$ increases by a factor of $\approx$~7 at the transition red$^{+}$$_{3}$, see \figref{fig:tunnel_spectroscopy}(b). This implies that for {\it V}$_{\text{probe}}$ from -0.5\,V to -0.1\,V, $\approx$~2/3 of the attached electrons create T$_{1,2}$ and $\approx$~1/3 create S$_0$, and at {\it V}$_{\text{probe}}$ from +0.2 V to +0.6 V, $\approx$~6/7 of the attached electrons create S$_{1,2}$ and $\approx$~1/7 create either S$_0$ or T$_{1,2}$. These observations indicate that when the voltage becomes large enough to access the next state higher in energy, the dominant fraction of transitions is into that state. This can be explained, as the NDR mentioned above, by a barrier-height effect~\cite{Repp2005Pentacene}: with increasing energy of the transition, the barrier height for the respective tunneling process becomes smaller, see \figref{fig:mechanism}(a-c). In addition, the ratios of the transitions will be affected by the local density of the respective orbitals (HOMO for red$^{+}$$_{1}$; LUMO and LUMO+1 for red$^{+}$$_{2}$ and red$^{+}$$_{3}$), and the spin multiplicity (number of channels) of the respective transitions. The barrier-height effect also explains the slight decrease of $I$, for {\it V}$_{\text{probe}}$ from -1.1\,V to -0.8\,V in \figref{fig:tunnel_spectroscopy}(a), as the barrier for resonant tunneling into the HOMO (red$^{+}$$_{1}$) increases with increasing voltages~\cite{Repp2005Pentacene, Wolkow}. The dominance of the conductance channel into the highest energetically allowed transition, which we observe and quantify here, has been hypothesized in light-emission STM experiments to explain the effective exciton formation in a double-barrier tunnel junction~\cite{Imada2016}.

Broadening of electronic states in tunneling spectroscopy of nanostructures on insulating films~\cite{Repp2005Cl} is linked to the ionic relaxation energy~\cite{marcus1993electron}. The differences in the measured FWHM of the transitions could be related to different relaxation energies, $\Delta_1$, $\Delta_2$ and $\Delta_3$ involved in each transition, see \figref{fig:mechanism}(d). However, the similar values of the calculated excited state energies in the NPc$^+$ and NPc$^0$ geometries (see supplemental material) suggest that the differences of the ionic relaxation energies $\Delta_1$, $\Delta_2$ and $\Delta_3$ are small. 
In addition to the relaxation energies, other effects will contribute to the FWHM, e.g. the NDR related to barrier-height effects will reduce the FWHM of the peaks and shift them to smaller energies. The limited energy resolution of the tunneling spectroscopy based on single-electron events and the large electron-phonon coupling on NaCl render it difficult to reveal vibronic sidepeaks~\cite{pavlivcek2013symmetry}. The latter have been resolved in single-electron charging experiments by AFM recently~\cite{roy2019fully}.

Our method complements the information obtained via light emission STM (LE-STM)~\cite{Qiu2003Science,Imada2016,Doppagne2018,zhang2016,zhang2017sub,Kuhnke2017,doppagne2020single}, which is usually performed in a double-barrier tunnel junction geometry. While our energy resolution is lower and the data acquisition time is larger compared to LE-STM, we can probe transitions into states that do not lead to light emission. Importantly, we can separately control and study the different electron detachment and attachment processes involved in the creation of an excited state. The latter allowed the quantification of the ratios of different transition channels. The method shown here does not require a double-barrier tunnel junction and should also be applicable on bulk insulators and semiconductor substrates with sufficiently large band gap, provided samples can be gated.

To conclude, we characterized the transitions to excited molecular states by AFM-based single-electron tunneling spectroscopy. We characterized three channels for the $\text{NPc}^{+}\rightarrow\text{NPc}^{0}$ transition. These include charge-state transitions between an out-of-equilibrium charged initial state ($\text{NPc}^{+}$) to out-of-equilibrium excited neutral final states that we interpret as triplet and singlet excited states. Depending on the voltage at which the reduction is performed, transitions to either of the assigned final states S$_{0}$, T$_{1,2}$ or S$_{1,2}$ occur dominantly. We quantified the excited state energies. 
The stepwise formation and probing of excited states introduced here could help in engineering energy transfer between molecules~\cite{Imada2016,zhang2016} and in unraveling mechanisms of molecular optoelectronic phenomena~\cite{Kuhnke2017}.

\begin{acknowledgments}
We thank J. Repp, D. Urbonas, K. Kaiser, Th. Frederiksen, P. Brandimarte, G. Meyer and R. Allenspach for discussions. We acknowledge the European Research Council (Consolidator grant 'AMSEL', agreement No. 682144) and the European FET-OPEN project SPRING (No. 863098) for financial support. 
\end{acknowledgments}

\bibliographystyle{apsrev4-1}
\input{Fatayer_ExcitedStates_AFM2020.bbl}

\end{document}

%% file: Fatayer_ExcitedStates_AFM2020.bbl
%